\documentclass[11pt]{article}
\usepackage[T1]{fontenc}
\usepackage[utf8]{inputenc}
\usepackage{mindbench-preprint}

\mindbenchstatus{Preprint v1.0.0}
\mindbenchdate{August 2026}
\mindbenchcode{https://github.com/mindbench-ai/search-source-audit}
\mindbenchdata{https://huggingface.co/datasets/MindBench/search-source-audit}
\mindbenchcorrespondence{jtorous@bidmc.harvard.edu}

\title{Sources of Truth: A Multi-Platform, Multilingual Audit of\\
Citations in AI Mental Health Information Queries}

\author{%
\begin{tabular}{c}
Phuong Anh Nguyen\textsuperscript{1} \quad Jill Noorily\textsuperscript{1} \quad
Matthew Flathers\textsuperscript{1} \quad Haruka Notsu\textsuperscript{1} \\[2pt]
Laura Ospina-Pinillos\textsuperscript{2} \quad Tommy Nguyen\textsuperscript{1,3} \quad
Samantha Clark\textsuperscript{1} \\[2pt]
Aoife Keane\textsuperscript{1} \quad Grace Thompson\textsuperscript{1} \quad
John Torous\textsuperscript{1}
\end{tabular}%
}

\mindbenchaffiliations{%
\textsuperscript{1}Division of Digital Psychiatry, Beth Israel Deaconess Medical Center,
Harvard Medical School, Boston, MA 02446, USA\par
\textsuperscript{2}Psychiatry and Mental Health Department, School of Medicine,
Pontificia Universidad Javeriana, Bogot\'a 110231, Colombia\par
\textsuperscript{3}Tufts University School of Medicine, Tufts University,
Boston, MA 02111, USA%
}

\mindbenchkeywords{generative AI; LLMs; online health information seeking; source credibility; information access; mental health; multilingual retrieval; web search audit; AI evaluation}

\begin{document}
\maketitle
\mindbenchresources

\begin{abstract}
Online health information seeking is shifting from keyword search, where users consider a ranked list of links, to conversational systems that compose a single answer and curate its citations. Source evaluation therefore passes from user to platform, yet what these systems surface is poorly characterized. We audited three free consumer products (ChatGPT, Perplexity, Google AI Overview) on twenty English mental health questions under two prompt conditions, with a subset of three also translated into six further languages of varying resource tiers. We recorded 15,942 citations across 1,140 responses and 1,713 unique domains, then classified every citation with a nine-category organizational typology applied by a deterministic classifier validated against human coding. Citations were heavily concentrated: the ten most-cited domains accounted for 43.6\% of English citations, and government, commercial health, and academic sources were closely matched at roughly 22\% each. Platforms differed little in typical citation volume but sharply in consistency and in the source types they favored. Explicitly requesting sources shifted composition only modestly. Non-English queries surfaced fewer citations and were routed to language-appropriate resources at significantly lower rates. We release the typology, classifier, and annotated corpus as reusable instruments for auditing generative health search.
\end{abstract}

\section{Introduction}\label{sec:introduction}

Over the past decade, Online Health Information Seeking (OHIS) has shifted from traditional keyword-based search engines toward conversational and generative AI interfaces such as ChatGPT (Fox \& Duggan, 2013; Wardle et al., 2025). On classic Search Engine Result Pages (SERPs), the work of information foraging falls to the user, who evaluates a ranked list of hyperlinks and judges each source for themselves (Pirolli \& Card, 1999). Conversational systems instead compose a single natural-language answer and curate the external citations that accompany it (Shah \& Bender, 2022), taking on the role of active information gatekeepers as some of the cognitive burden of source triage passes from the user to the platform (Metzger \& Flanagin, 2013). Because the results these systems present are likely to be more trusted than ranked links (Metzger et al., 2010; Yun \& Bickmore, 2025), identifying which sources they deliver to end users is an important object of research.

This change in health information foraging behavior happens to coincide with increasing user reliance on AI for mental health guidance, and for support more broadly (Bodner et al., 2026; McBain et al., 2025b). Mental health distress continues to rise, a trend particularly evident in the years following the COVID-19 pandemic, and this sustained burden has driven a corresponding increase in online searches for mental health information and support (Du et al., 2020). People who search for mental health information online are often motivated by concerns over personal privacy and an immediate need for accessible guidance (Pretorius et al., 2019), and these motivations have made conversational AI an increasingly common entry point for psychoeducation and symptom navigation (Li et al., 2023; Bodner et al., 2026).

However, what these systems surface is poorly characterized. Evaluation of AI and mental health has concentrated on high-acuity crisis interventions, suicide-risk detection and self-harm guardrails (Flathers et al., 2026; McBain et al., 2025a; Arnaiz-Rodriguez et al., 2026). This is valuable safety work, but mental health queries span a much wider spectrum, from everyday distress and subclinical concerns to severe, acute crises, and the majority of user queries involve routine, non-crisis psychoeducation (Stade et al., 2026; McBain et al., 2025b). Health-domain evaluation is also model-centric, concentrating on accuracy in medical-examination question answering rather than real-world use (Rauh et al., 2024; Bedi et al., 2025; Hua et al., 2025). Users, however, encounter products rather than models: a model wrapped in additional layers such as web search and citation interfaces. Source selection is impacted by that wrapper, and the audit literature that does exist has found generated citations that fail to support the text they accompany, and retrieval behavior that varies across repeated executions of the same query (Liu et al., 2023; Narayanan Venkit et al., 2025; Li \& Sinnamon, 2024; Kirsten et al., 2026).

Finally, AI evaluation literature remains disproportionately anglocentric, masking severe cross-lingual disparities in health information routing (Blasi et al., 2022; Lai et al., 2023; Kuai et al., 2026). Generative search in non-English contexts operates under a dual scarcity constraint: less authoritative native-language health content exists on the open web, and cross-lingual retrieval fidelity is lower (Blasi et al., 2022; Joshi et al., 2020; Ahuja et al., 2023). Systems may therefore retrieve English-language evidence even for queries submitted in lower-resource languages, returning answers translated into the user\textquotesingle s language while citing the original English sources (Park et al., 2026; Ranaldi et al., 2026), with elevated risk of broken citations or an over-reliance on machine-translated Western sources that ignore local clinical paradigms (Magueresse et al., 2020; Kuai et al., 2026).

To our knowledge, no prior work has audited mental health queries specifically, and few have examined how ordinary variation in how users ask, such as explicitly requesting sources, changes which sources appear. This study provides an empirical audit of three free consumer-facing AI search products currently acting as gatekeepers of mental health information, addressing four dimensions of how they surface sources: (1) source characteristics and classification, for which we develop and release an organizational typology, with the annotated corpus and the deterministic classifier that produced it, as reusable instruments for auditing generative health search; (2) analysis of citation and source-routing patterns across common conversational platforms; (3) the effect of prompting for source attribution on citation behavior; and (4) citation patterns across different language contexts.

\section{Methods}\label{sec:methods}

\subsection{Study design and overview}\label{sec:study-design-and-overview}

We conducted an empirical audit of three freely accessible, consumer-facing AI search products (ChatGPT, Perplexity, Google's AI Overview) to characterize the external sources surfaced in response to mental health information-seeking queries. The audit varied four factors: platform (3), query language (7), prompt condition (2; with vs.~without an explicit request for sources), and question (20 in English and a consistent subset of 3 in each non-English language). Each condition was administered five times by a team of seven trained annotators, who recorded every cited external source. The primary outcomes were the number and identity of sources surfaced under each condition. All product-level data was collected between June 25th 2026 and July 20th 2026.

We audited the consumer product because it is the user-facing surface. However, the product is not equivalent to the underlying model: its responses reflect both the model and additional systems that retrieve, select, and cite sources. We therefore conducted a parallel audit of comparable models accessed via the three vendors' developer APIs, run on July 23rd, 2026 and reported in Appendix S1.

\subsection{Question set development}\label{sec:question-set-development}

We developed a set of 20 English-language questions reflecting routine mental health information-seeking behaviors. The questions were organized into eight non-exclusive functional categories: signs and symptoms (n = 4; e.g., ``What are the symptoms of bipolar disorder?''), diagnosis (n = 3; e.g., ``How is PTSD diagnosed?''), treatment effectiveness (n = 3; e.g., ``How effective is exposure therapy for OCD?''), medications (n = 3; e.g., ``What medications are available for depression?''), medication side effects (n = 1; ``What are the side effects of SSRIs?''), treatment guidelines (n = 2; e.g., ``How is PTSD treated according to clinical guidelines?''), crisis resources (n = 3; e.g., ``What crisis resources are there for psychosis?''), and identification of trustworthy information sources (n = 1; ``What are some of the most trustworthy sources of mental health information?''). The question set was designed to capture a range of psychoeducational queries that reflect likely domains of real-world mental health search activity. The complete list of questions is provided in Appendix S2.

\subsection{Platform selection}\label{sec:platform-selection}

The three analyzed platforms were selected for their prominence among consumer AI search products and free tier limits sufficient for the scope of this experiment. For each platform, we used the free, default configuration a typical user would encounter: ChatGPT and Perplexity in their base free tiers, and Google\textquotesingle s AI Overview accessed by entering the query into Google Search. To limit personalization and history effects, annotators were instructed to query each platform while logged out, in a fresh or cleared session, using incognito/private browsing where possible. The full annotator instructions are in Appendix S3.

\subsection{Prompt conditions}\label{sec:prompt-conditions}

Each question was administered under two prompt conditions to assess whether an explicit request for sources altered citation behavior: (1) a question-only condition, representing the default conversational format, in which the question was submitted verbatim; and (2) a question-plus-source-request condition, in which the instruction ``List your Sources.'' was appended to the question.

\subsection{Language selection and translation}\label{sec:language-selection-and-translation}

To assess cross-lingual disparities, we administered queries in seven languages spanning three resource tiers, operationalized following established taxonomies of digital language-resource availability (Joshi et al., 2020): high-resource (English, Spanish, Japanese), medium-resource (Ukrainian, Hindi), and low-resource (Nepali, Twi). Because a full translation set of all 20 questions was beyond the scope of the volunteer effort, we translated three questions into each non-English language, selected to represent functionally distinct query types: symptom/diagnostic (``How is depression diagnosed?''), acute crisis (``What crisis resources are there for suicidal thoughts?''), and source-seeking meta-information (``What are the most trustworthy sources of mental health information?''). Translations were produced by native speakers of each language. The complete list of translations is provided in Appendix S2.

\subsection{Source extraction}\label{sec:source-extraction}

Annotators recorded sources from every citation channel presented in a response, including in-text citations, clickable source links, and end-of-response reference lists. We then parsed the annotations programmatically to extract each citation\textquotesingle s destination URL. URLs were standardized to their registrable domain (the core domain that can be registered independently, e.g., nih.gov) using the Public Suffix List (Mozilla Foundation, 2026). We then derived the host (e.g., www.nimh.nih.gov), top-level domain, URL validity, and two complementary indicators of language-appropriate routing: whether the citation sat on a country-code top-level domain (ccTLD) associated with the query language (e.g., .jp for Japanese), and whether its URL carried a native-language signal such as a path locale code, percent-encoded native script, or explicit language parameter. The second captures native-language pages served from non-local domains (e.g., medlineplus.gov/spanish/) that the ccTLD measure alone misses. The resulting dataset is released as an annotated corpus at \url{https://huggingface.co/datasets/MindBench/search-source-audit}.

\subsection{Source-type typology and annotation}\label{sec:source-type-typology-and-annotation}

To characterize what kinds of sources platforms route users toward, we developed a nine-category organizational typology and applied it to every citation via an ordered, rule-based classifier operating on the registrable domain, the host, and the public suffix:

\begin{quote}
1. Government / public health: .gov and country-government domains, plus intergovernmental (WHO) and national health-service (NHS) sites;

2. Academic / journal: clinical literature (PubMed, PMC, NCBI), universities (.edu), and scholarly publishers or DOIs;

3. Nonprofit health system: patient-education content from nonprofit and university-affiliated hospitals and health systems (e.g., Mayo Clinic, Cleveland Clinic, Johns Hopkins);

4. Commercial health: for-profit health-information publishers and telehealth services, pharmaceutical and branded-drug sites, private hospital groups, and private clinical practices (e.g., WebMD, Healthline, GoodRx, BetterHelp);

5. Nonprofit / advocacy: non-governmental organizations and advocacy groups (e.g., NAMI, the American Psychiatric Association);

6. Encyclopedia (wiki): Wikipedia and comparable reference wikis;

7. News / media;

8. Social / video: social platforms, forums, and video (e.g., YouTube, Reddit, Facebook);

9. Other.
\end{quote}

After algorithmic classification, the assignments were validated against human coding in two stages. First, the 100 most-cited domains, which account for the majority of citations, were reviewed and finalized in full by two authors. Second, a random sample of 200 domains in the remaining set, distributed as evenly as possible across the nine categories, was reviewed independently by two annotators.

\section{Results}\label{sec:results}

\subsection{Overview}\label{sec:overview}

We collected 1,140 responses, totaling 15,942 recorded citations across 1,713 unique domains. The English-language subset comprised 600 responses containing 10,972 citations to 1,038 unique domains. The non-English subset comprised 540 responses (90 per language) containing 4,970 citations to 870 unique domains. Including all responses (n =1,140), 83 responses (7.3\%) contained no citations. No citations in responses were most common for ChatGPT (55/380, 14.5\%), followed by Google AI Overview (27/380, 7.1\%) and Perplexity (1/380, 0.3\%).

\subsection{Citation volume}\label{sec:citation-volume}

Across the 600 English-language responses (n = 200 per platform), mean citation counts differed sharply while medians did not (Figure 1A). Medians clustered at 10-12 for all three products, but ChatGPT\textquotesingle s mean was 32.29 with an SD of 60.50, against 12.85 (SD 6.63) for Google AI Overview and 9.7 (SD 1.69) for Perplexity. Platform differences in volume were therefore driven by a subset of outlier responses, almost all from ChatGPT.

Citation count also varied by question category, from a mean of 44.2 for medication questions to 9.7 for crisis resources (Figure 1B). Medians again clustered near 10, and the ten largest responses in the corpus were all produced by ChatGPT, all answering medication or treatment-guideline questions.

\begin{figure}[H]
  \centering
  \includegraphics[width=\linewidth]{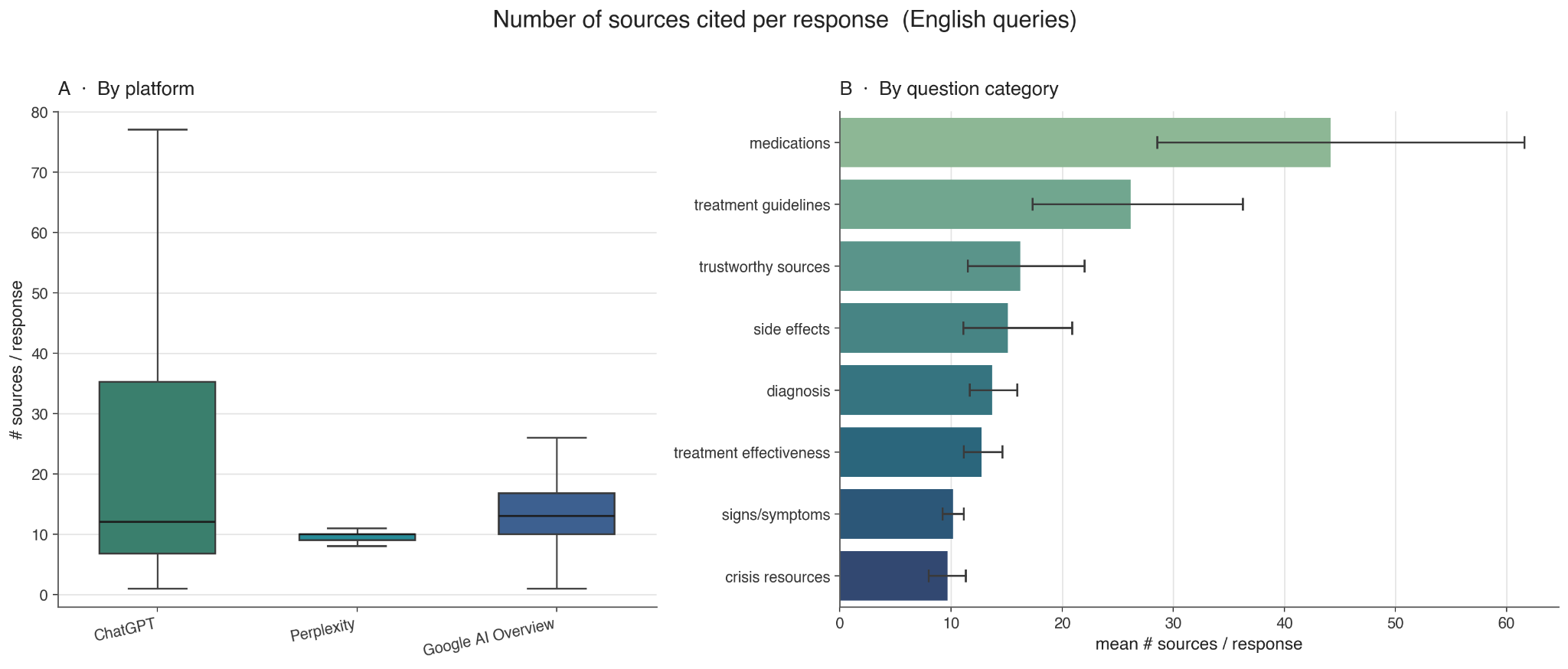}
  \caption{Citation volume by platform and question type. Figure 1A shows the number of sources per response (English query only) for ChatGPT, Perplexity and Google AI Overview; ChatGPT had the highest mean but similar medians to the others. Figure 1B breaks down citation counts by question category, showing the mean number of sources per response.}
  \label{fig:1}
\end{figure}

\subsection{Citation Source Types}\label{sec:citation-source-types}

Classifier assignments have high agreement with independent human coding. On the 200-domain validation sample, Cohen\textquotesingle s κ was 0.88 between the two annotators, and 0.89 and 0.86 between each annotator and the classifier.

Of all the English-question citations, government sources and commercial health were the most cited source types (22.6\% each), followed by academic/journals (21.6\%), nonprofit/advocacy (13.3\%), nonprofit health systems (11.6\%), encyclopedia (3.9\%), social/video (3.0\%), unclassified other (0.7\%), and news media (0.7\%) (Figure 2). The citations presented for English-questions were heavily concentrated to a small array of domains (Figure 3). The top ten most cited sources accounted for 43.6\% of all 10,959 English citations.

\begin{figure}[H]
  \centering
  \includegraphics[width=\linewidth]{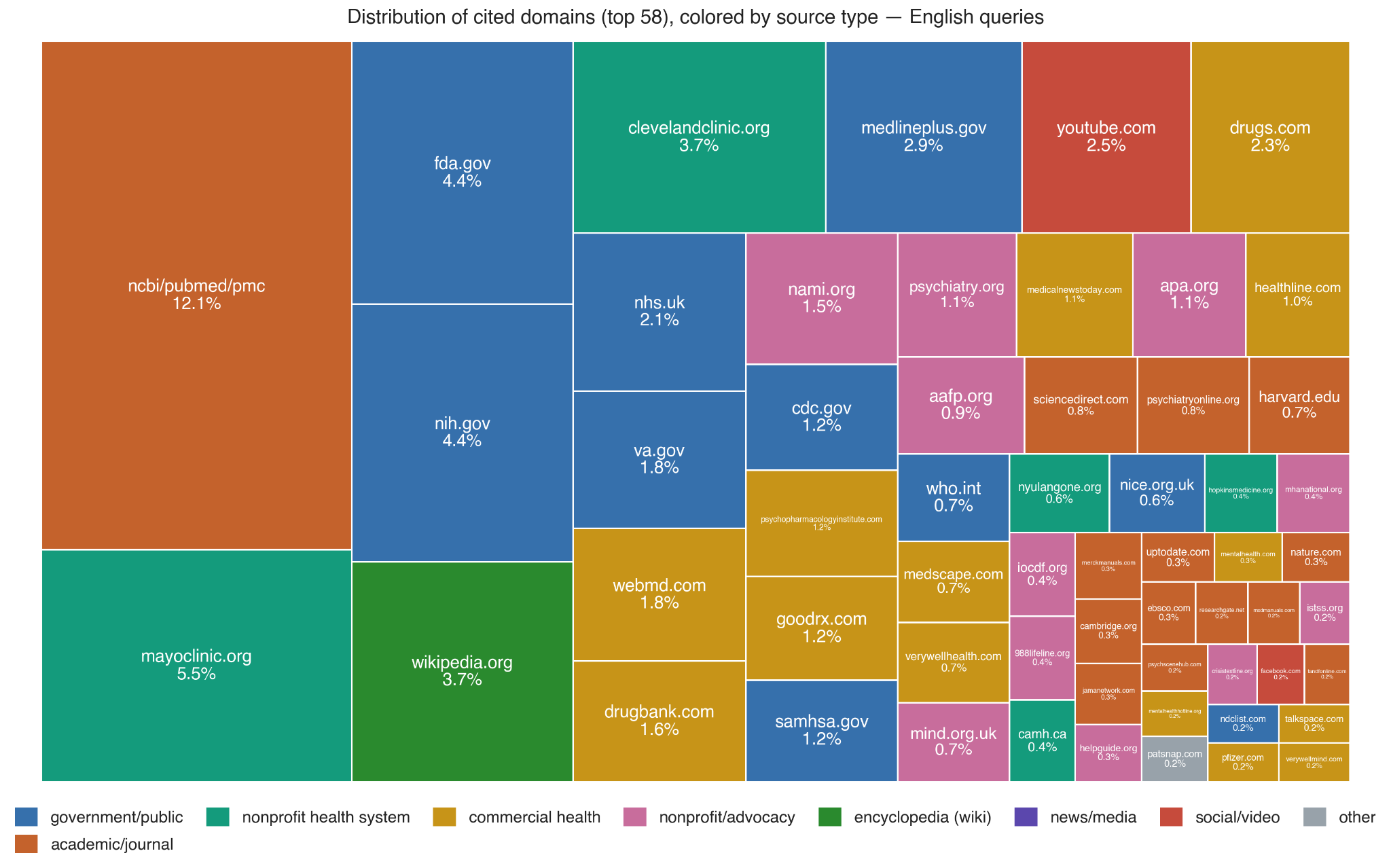}
  \caption{A distribution chart of the top 58 domains cited across English-language queries, color-coded by source type (academic, government, commercial health, etc.), showing how citations concentrate heavily on a small handful of sites like PubMed/NCBI and Mayo Clinic.}
  \label{fig:2}
\end{figure}

Platforms differed in source-type composition (Figure 3A). Each led with a different type: government/public for ChatGPT (26.4\%), academic/journal for Perplexity (24.3\%), and commercial health for Google AI Overview (25.7\%). The sharpest contrasts were in the smaller categories: ChatGPT cited Wikipedia far more than the others (6.1\% vs ≤0.5\%), while Google AI Overview cited social/video far more (8.2\% vs ≤1.5\%), mostly YouTube (7.7\%). News/media was negligible for all three (≤1.0\%).

Source type also tracked question category (Figure 3B). The pairings were largely intuitive: treatment-effectiveness questions drew most heavily on academic/journal sources (56.9\%) and crisis questions on nonprofit/advocacy (41.6\%), with one exception of note: medication questions drew the largest commercial-health share of any category (33.7\%).

\begin{figure}[H]
  \centering
  \includegraphics[width=\linewidth]{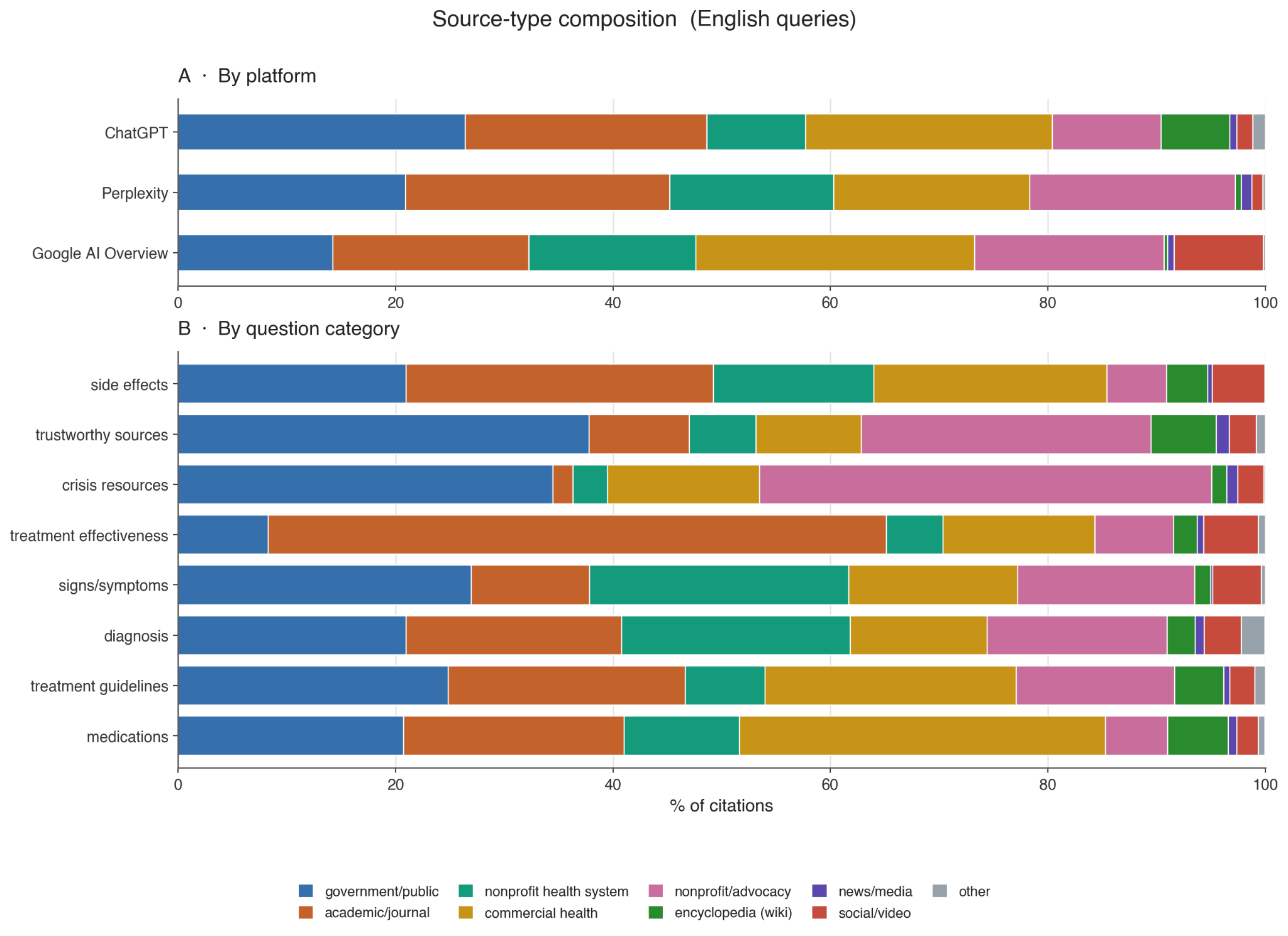}
  \caption{Source-type composition for English queries. Figure 3A shows the percentage of citation types by platform. Figure 3B shows the percentage of citation types by question category.}
  \label{fig:3}
\end{figure}

\subsection{Effect of explicitly requesting sources}\label{sec:effect-of-explicitly-requesting-sources}

Appending "List your Sources." had little effect on the overall number of citations surfaced: across all English responses, the mean rose only modestly from 16.6 in the question-only condition to 19.9 when sources were explicitly requested (Figure 4A). This shift was driven almost entirely by ChatGPT, which showed a more substantial change (mean 27.3 to 37.2 citations). Perplexity (mean 9.2 to 10.3) and Google AI Overview (mean 13.4 to 12.3) surfaced essentially the same number of sources whether or not sources were explicitly requested.

Requesting sources was associated with small shifts in source composition (Figure 4B), with slightly larger proportions of government/public (+2.1 percentage points), commercial health (+1.6), and academic/journal (+0.8) citations, and slightly smaller proportions of nonprofit health system (-1.6), nonprofit/advocacy (-1.3), social/video (-1.0), and other (-0.4) citations.

\begin{figure}[H]
  \centering
  \includegraphics[width=\linewidth]{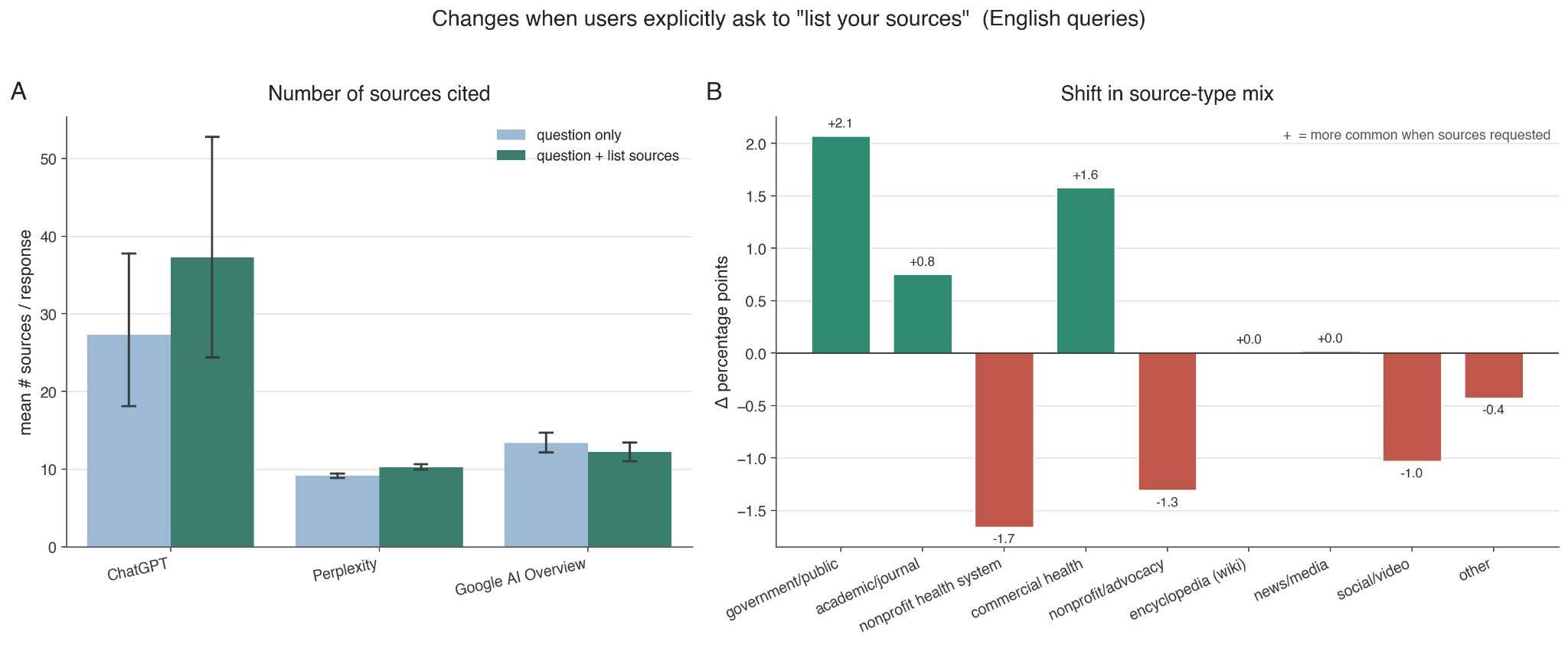}
  \caption{Effect of explicitly asking for sources. Figure 4A shows the shift in citation volume between question-only and ``List your Sources'' conditions across platforms. Figure 4B shows the resulting small shift in source-type composition when ``List your Sources'' is appended.}
  \label{fig:4}
\end{figure}

\subsection{Cross-lingual citation behavior}\label{sec:cross-lingual-citation-behavior}

On the three questions administered in all seven languages, non-English queries surfaced fewer sources than English ones (mean 9.2, SD 8.7, n=540 vs 12.3, SD 11.6, n=90), ranging from 11.2 for Spanish to 7.1 for Twi (Figure 5A). The gap was driven by the products that fail rather than by a uniform decline: ChatGPT averaged 17.5 sources in English against 4.3 in Twi and returned none for 33\% of Twi and 23\% of Japanese responses, and Google AI Overview returned none for 30\% of Hindi responses, while Perplexity stayed within 8.6-10.2 across all seven languages.

For non-English queries, the degree to which AI systems directed users to language-appropriate resources varied substantially by language (Figure 5B). When localization was measured only by citations to ccTLDs, high-resource languages (Japanese, Spanish) received 38.0\% of citations on national domains, compared with 20.8\% for medium-resource (Ukrainian, Hindi) and just 3.3\% for low-resource languages (Twi, Nepali). However, this measure substantially underestimated localization because many citations pointed to translated content hosted on international domains rather than country-specific websites. Spanish-language queries, for example, were most often directed to Spanish-language pages on United States government domains, such as the Spanish section of MedlinePlus (\href{http://medlineplus.gov/spanish/}{\ul{medlineplus.gov/spanish/}}). Counting native-language pages on non-local domains raised the language-appropriate share for every language, most dramatically for Twi (2.5\% to 37.6\%) and Spanish (11.3\% to 46.2\%) (Figure 5B).

\begin{figure}[H]
  \centering
  \includegraphics[width=\linewidth]{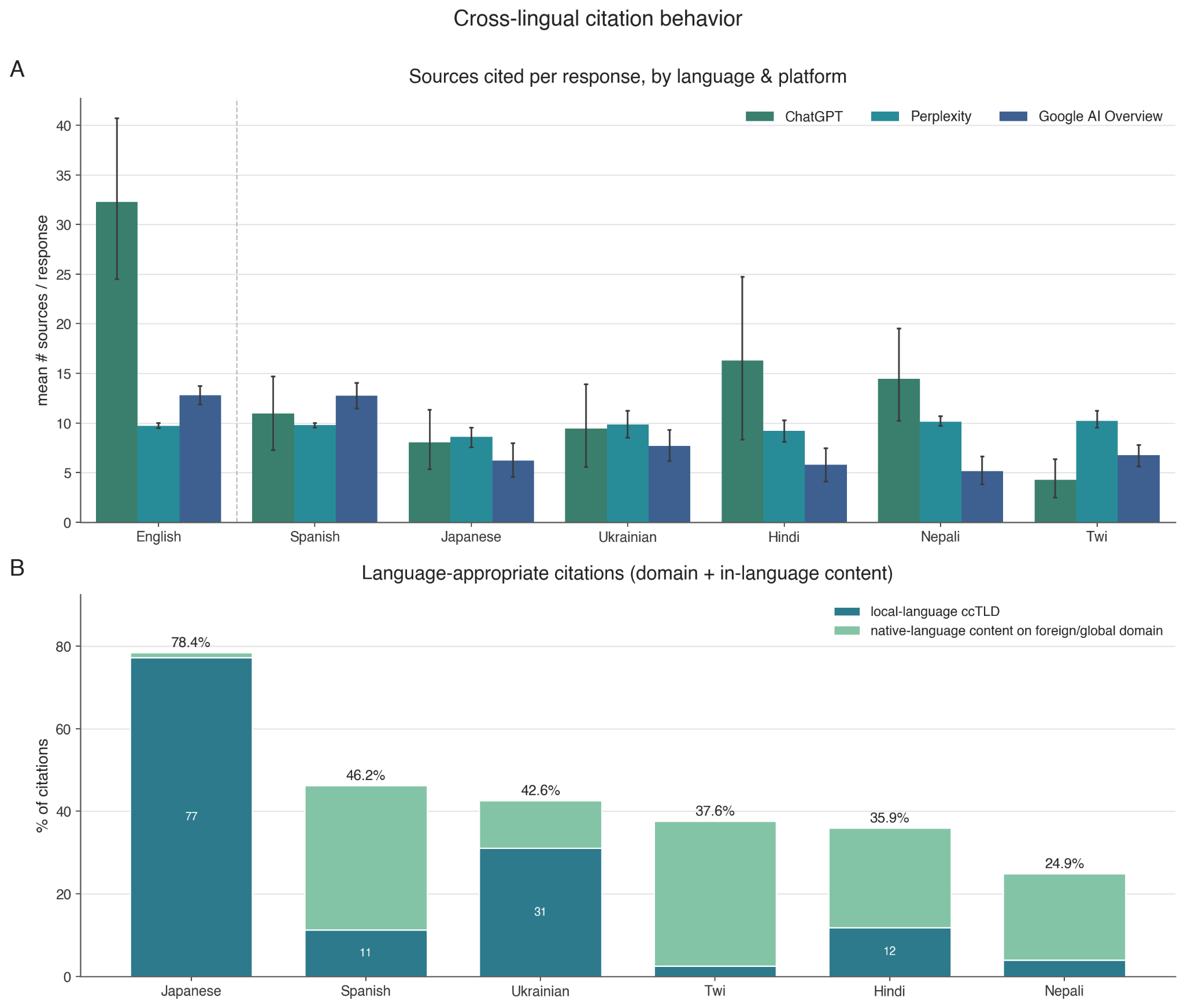}
  \caption{Cross-lingual results. Figure 5A compares citation volume across platforms and languages. Figure 5B shows the share of citations pointing to language-appropriate resources (ccTLD or native-language).}
  \label{fig:5}
\end{figure}

\subsection{Sensitivity to individual annotators}\label{sec:sensitivity-to-individual-annotators}

As a sensitivity analysis covering all languages, every reported estimate was recomputed seven times, each time excluding one annotator\textquotesingle s contributions (leave-one-annotator-out jackknife). Across 82 estimates, the median largest absolute deviation from the full-sample value was 0.82 percentage points (90th percentile 3.25). Every platform-level and question-category ranking held under all seven exclusions. The two least stable estimates were ChatGPT\textquotesingle s mean English citation count (range 24.1--38.9) and ChatGPT\textquotesingle s no-citation rate (8.2--17.1\%), both reflecting a small number of responses with unusually large or empty citation lists. Per-estimate results are provided in Appendix S4.

\section{Discussion}\label{sec:discussion}

This study examined the citations that three consumer AI products (ChatGPT, Perplexity, and Google AI Overview) provide users when answering mental health-related questions. We identified four primary findings: citations concentrated on a small group of institutional sources; platforms differed little in median citation count but sharply in consistency and source-type mix; explicitly asking for sources changed little; and citation behavior deteriorated for many non-English queries.

AI search engines and chatbots now act as gatekeepers, determining which webpages users are directed to when seeking mental health information. But we found that all 3 products concentrate users\textquotesingle{} attention on a narrow set of citations. While these institutional sources are broadly authoritative, channeling users through such a narrow set of gatekept resources also concentrates risk -- a handful of pages effectively define the information environment for sensitive health questions, so any gaps, errors, or biases they contain propagate to users at scale. In addition, the type of question slightly changed which type of sources were presented, shifting this risk. For instance, medication questions, where commercial incentives are arguably the strongest, relied most heavily on commercial health sources, mainly routing users to sites such as \ul{drugs.com} and GoodRx.

Although the three platforms differed little in citation volume (median counts clustered at 10-12 sources per response), they differed sharply in consistency. This likely reflects a structural difference in how each product surfaces sources: Perplexity presents a fixed number of source cards, yielding near-constant counts, while ChatGPT embeds hyperlinks inline throughout its generated answer with no apparent cap, so citation volume can balloon. This variability produces an unpredictable user experience in which similar questions can return just a handful or up to a hundred sources.

Platforms also differed in the types of sources they favored, information that is useful to users choosing among products. All three drew primarily on academic, government, commercial, and nonprofit health system sources, but with distinct tendencies: Google AI Overview leaned toward commercial health and social/video, ChatGPT toward encyclopedias, and Perplexity toward nonprofit and advocacy sources. The platform a user selects can therefore materially shape the mental health information they receive.

When searching for health information, a common strategy some users use to get more valid information is to explicitly ask the chatbots to ``list their sources''. Our results show that taking that extra step yields little benefit: Citation volume was essentially unchanged and source type composition shifted only modestly (by at most 2.1 percentage points, in a more institutional direction, more towards government and academic sources).

\subsection{Cross-lingual citations landscape}\label{sec:cross-lingual-citations-landscape}

One concern in AI research is that models often privilege English-language queries, which can widen existing cross-lingual disparities (Rohera et al., 2025). Our findings support this concern and show that English-language citations are consistently prioritized. For instance, for Spanish, only 11.3\% of their citations were from countries in which Spanish is formally a national language such as Spain and Mexico.

These findings suggest that conversational AI retrieves the most visible and indexable language-specific resources available on the public web rather than preferentially routing users to health systems within countries where a language is spoken. The pattern across languages follows the supply of authoritative in-language material rather than the language\textquotesingle s resource tier. Japanese, the best-localized language in our data, has an extensive domestic institutional web; Spanish localization runs almost entirely through translated pages on United States government domains; and Ukrainian draws heavily on foreign government resources produced for refugee populations. This framing, however, risks equating language with cultural and contextual relevance. For instance, Spanish is the primary language across more than twenty countries, yet Spanish-language queries may disproportionately surface English-to-Spanish translations of U.S. institutional sources such as the NIH -- content that is likely well-suited to Spanish-speaking users within the United States but not necessarily to users in other Spanish-speaking countries, who may be more familiar with, and better served by, their own countries\textquotesingle{} national health authorities or locally recognized institutions. Improving multilingual mental health information access will therefore require more than advances in multilingual retrieval. It will also require continued investment in authoritative, publicly accessible mental health resources across diverse languages and greater attention to how retrieval systems prioritize local versus international sources.

\subsection{Anomalies}\label{sec:anomalies}

In a few cases, seemingly irrelevant links were presented to queries instead of mental health citations. Upon inspection, anomalous citations typically followed a similar pattern: they were all lexical neighbors of other more legitimate citations. For instance, a query seeking information about anxiety medications was directed to Wikipedia's entry on mayonnaise, alongside a dozen Mayo Clinic links. A query asking how depression is diagnosed returned results for the fragrance conglomerate DSM-Firmenich and the storage operating system Synology DSM, potentially because they share the acronym DSM with the \emph{Diagnostic and Statistical Manual of Mental Disorders}, the clinical reference used to diagnose depression. A query about PTSD surfaced a construction company sharing the acronym PCL-5 with the PTSD Checklist for DSM-5, a standard PTSD screening instrument. Similarly, a Hindi-language query asking how depression is diagnosed was shown Broadway tickets to \emph{Hamilton} alongside results for the Hamilton Depression Rating Scale (HAM-D). These anomalies connect to a documented class of cybersecurity vulnerability. Bad actors routinely register domains lexically adjacent to trusted institutional names (Kintis et al., 2017), through practices such as ``typosquatting'' (misspellings) and ``combosquatting'' (a brand name combined with another word). Retrieval behavior that mistakes mayonnaise for the Mayo Clinic suggests a level of matching on lexical proximity, and lexical proximity is occupiable: an adversarial attack could seed health misinformation, phishing infrastructure, or product advertising in pages and domains adjacent to the very narrow set of most commonly retrieved clinical targets and wait for retrieval systems to pull it into answers. We did not find any lexical neighbor serving misleading health content in our dataset, but the same similarity that produced these harmless errors could be deliberately exploited in the future.

\subsection{Clinical implications for the help-seeking process}\label{sec:clinical-implications-for-the-help-seeking-process}

These findings have important implications for an early and often vulnerable stage of mental health help-seeking, when individuals turn to conversational AI before or instead of consulting a clinician. Such platforms can improve access to understandable and immediate psychoeducation and generally direct English-language users toward a legitimate albeit very narrow set of government, academic, and health-system sources. However, this benefit is accompanied by substantial and largely invisible variation in citation quality and availability across platforms, question types, and languages. Because users may be unable to recognize these limitations from the tone or presentation of an AI response, digital literacy and regulation should be viewed as complementary safeguards: helping users identify poorly supported or contextually inappropriate information while establishing expectations for citation transparency, consistency, and equitable performance across languages. Our findings do not argue against conversational AI in mental health help-seeking, but rather for treating it as a useful yet imperfect tool that requires supporting structures to reduce the consequences of its uneven performance and heightened perceived trustworthiness as an information gatekeeper.

\subsection{Limitations}\label{sec:limitations}

This is a single-point audit (June/July 2026) of rapidly evolving products. Retrieval behavior might change and our estimates are only a snapshot of product behaviors. We also audited only free tiers in logged out sessions to limit personalization. Behavior for logged-in, paying tier customers may differ. Working in the free tier level limits our ability to record the specific models used for each product, which is an issue of reproducibility that comes with product level experimentation. Our link validity measure captures syntactic well-formedness only; we did not verify that links resolved or cited the correct pages. Only three questions were administered per non-English language, so non-English estimates are based on a narrower query base than English estimates. Although the translations were produced by native speakers, they do introduce possibilities of phrasing variations. In addition, all of our annotators were based in the USA, which is visible through their IP address and related metadata and may have impacted the volume at which international links were pulled or localized links were sourced.

\section{Conclusion}\label{sec:conclusion}

Consumer AI products have quietly become a primary triage point for mental-health information seeking, and our audit shows that the sources they surface are neither neutral nor uniform. For English-language users, all three products concentrated attention on a small institutional core: roughly three-quarters of citations came from academic, government, commercial-medical, and nonprofit health-system sources, and ten destinations alone accounted for 43.6\% of them. Each product also led with a different source type, so users' product choices shape the evidence they are shown. For non-English users, what localization existed tracked the supply of authoritative in-language material rather than any property of the language itself, so the binding constraint lies in what has been published and indexed as much as in what these systems can retrieve. We release the nine-category typology, the deterministic classifier, and the annotated corpus of 15,929 classified citations so that the source composition of health queries search can be measured comparably across systems, languages, and over time. Ensuring equitable access to trustworthy mental-health information in the AI search era will require routine product-level audits of cited sources, sustained investment in authoritative and publicly accessible multilingual mental-health content, and greater attention from public-health agencies and AI companies to the reach and language coverage of the web resources on which these systems depend.

\section{Generative AI Disclosure}\label{generative-ai-disclosure}

Claude Opus 5 aided with the initial drafting of the rule-based classifier and with copyediting of the main manuscript. Claude Fable 5 aided with the drafting of Appendix S1. In all cases, the output was reviewed, corrected, and finalized by the authors. No generative AI was used to create or modify the study data.

\section{Conflict of Interest Statement}\label{conflict-of-interest-statement}

The authors declare no conflicts of interest.

\section{Funding Statement}\label{funding-statement}

This research received no specific grant from any funding agency in the public, commercial, or not-for-profit sectors.

\section{Data Availability Statement}\label{data-availability-statement}

The data that support the findings of this study are openly available on Hugging Face at \url{https://huggingface.co/datasets/MindBench/search-source-audit}, DOI 10.57967/hf/10067. The source-type classifier and the analysis code that generated all reported estimates and figures are openly available at \url{https://github.com/mindbench-ai/search-source-audit}.

\section{Ethics Statement}\label{ethics-statement}

This study did not involve human participants or identifiable human-subject data. The annotators acted as data collectors rather than as research participants. The study was therefore not human-subjects research and did not require institutional review board approval.

\section{Acknowledgments}\label{sec:acknowledgments}

The authors thank the additional volunteers who administered queries and recorded source citations across platforms and languages: Valerie Rosario, Anna Simms, and Mihika Hete. We also thank the translators who contributed the non-English question sets: Eva Donahue, Svitlana Bielichenko, Antara Tripathy, Asmi Shrestha, Glenn Asuo-Asante, and Lordina Amoako.

Laura Ospina-Pinillos gratefully acknowledges the support of The Fulbright Program and the Fulbright Colombia Commission, which supported her Visiting Scholar appointment at the Division of Digital Psychiatry, Beth Israel Deaconess Medical Center, Harvard Medical School.

\section*{References}
\begingroup
\setlength{\parindent}{0pt}
\setlength{\parskip}{6pt}
\everypar{\hangindent=0.5in\hangafter=1}

Ahuja, K., Diddee, H., Hada, R., Ochieng, M., Ramesh, K., Jain, P., Nambi, A., Ganu, T., Segal, S., Axmed, M., Bali, K., \& Sitaram, S. (2023). MEGA: Multilingual evaluation of generative AI. In Proceedings of the 2023 Conference on Empirical Methods in Natural Language Processing (pp. 4232--4267). Association for Computational Linguistics. \url{https://doi.org/10.18653/v1/2023.emnlp-main.258}

Arnaiz-Rodriguez, A., Baidal, M., Derner, E., Layton Annable, J., Ball, M., Ince, M., Perez Vallejos, E., \& Oliver, N. (2026). Between help and harm: An evaluation study of mental health crisis handling by large language models. JMIR Mental Health, 13, e88435. \url{https://doi.org/10.2196/88435}

Bedi, S., Liu, Y., Orr-Ewing, L., Dash, D., Koyejo, S., Callahan, A., Fries, J. A., Wornow, M., Swaminathan, A., Lehmann, L. S., Hong, H. J., Kashyap, M., Chaurasia, A. R., Shah, N. R., Singh, K., Tazbaz, T., Milstein, A., Pfeffer, M. A., \& Shah, N. H. (2025). Testing and evaluation of health care applications of large language models: A systematic review. JAMA, 333(4), 319--328. \url{https://doi.org/10.1001/jama.2024.21700}

Blasi, D., Anastasopoulos, A., \& Neubig, G. (2022). Systematic inequalities in language technology performance across the world's languages. In Proceedings of the 60th Annual Meeting of the Association for Computational Linguistics (Volume 1: Long Papers) (pp. 5486--5505). Association for Computational Linguistics. \url{https://doi.org/10.18653/v1/2022.acl-long.376}

Bodner, R., Lim, K., Siddals, S., Goldberg, S., \& Torous, J. (2026). Barriers to understanding how many people use AI for mental health support: An estimate and narrative review. npj Digital Public Health, 1, Article 21. \url{https://doi.org/10.1038/s44482-026-00025-7}

Du, H., Yang, J., King, R. B., Yang, L., \& Chi, P. (2020). COVID-19 increases online searches for emotional and health-related terms. Applied Psychology: Health and Well-Being, 12(4), 1039--1053. \url{https://doi.org/10.1111/aphw.12237}

Flathers, M., Nguyen, P. A. H., Noorily, J., Herpertz, J., Cong, S., Calvert, E., Oh, A., Li, S., Lane, E., Ledley, K., Bodner, R., \& Torous, J. (2026). Mental health benchmarks for large language models: A systematic scoping review [Preprint]. PsyArXiv. \url{https://doi.org/10.31234/osf.io/fahuk_v1}

Fox, S., \& Duggan, M. (2013). Health online 2013. Pew Research Center's Internet \& American Life Project. \url{https://www.pewinternet.org/wp-content/uploads/sites/9/media/Files/Reports/PIP_HealthOnline.pdf}

Hua, Y., Siddals, S., Ma, Z., Galatzer-Levy, I., Xia, W., Hau, C., Na, H., Flathers, M., Linardon, J., Ayubcha, C., \& Torous, J. (2025). Charting the evolution of artificial intelligence mental health chatbots from rule-based systems to large language models: A systematic review. World Psychiatry, 24(3), 383--394. \url{https://doi.org/10.1002/wps.21352}

Joshi, P., Santy, S., Budhiraja, A., Bali, K., \& Choudhury, M. (2020). The state and fate of linguistic diversity and inclusion in the NLP world. In Proceedings of the 58th Annual Meeting of the Association for Computational Linguistics (pp. 6282--6293). Association for Computational Linguistics. \url{https://doi.org/10.18653/v1/2020.acl-main.560}

Kintis, P., Miramirkhani, N., Lever, C., Chen, Y., Romero-Gómez, R., Pitropakis, N., Nikiforakis, N., \& Antonakakis, M. (2017). Hiding in plain sight: A longitudinal study of combosquatting abuse. In Proceedings of the 2017 ACM SIGSAC Conference on Computer and Communications Security (pp. 569--586). Association for Computing Machinery. \url{https://doi.org/10.1145/3133956.3134002}

Kirsten, E., Große Perdekamp, J., Wu, Q., Upadhyay, M., Gummadi, K. P., \& Zafar, M. B. (2026). Characterizing web search in the age of generative AI. In Findings of the Association for Computational Linguistics: ACL 2026 (pp. 10827--10848). Association for Computational Linguistics. \url{https://doi.org/10.18653/v1/2026.findings-acl.526}

Kuai, J., Brantner, C., Karlsson, M., Van Couvering, E., \& Romano, S. (2026). AI chatbot accountability in the age of algorithmic gatekeeping: Comparing generative search engine political information retrieval across five languages. New Media \& Society, 28(5), 2121--2143. \url{https://doi.org/10.1177/14614448251321162}

Lai, V. D., Ngo, N. T., Pouran Ben Veyseh, A., Man, H., Dernoncourt, F., Bui, T., \& Nguyen, T. H. (2023). ChatGPT beyond English: Towards a comprehensive evaluation of large language models in multilingual learning. In Findings of the Association for Computational Linguistics: EMNLP 2023 (pp. 13171--13189). Association for Computational Linguistics. \url{https://doi.org/10.18653/v1/2023.findings-emnlp.878}

Li, A., \& Sinnamon, L. (2024). Generative AI search engines as arbiters of public knowledge: An audit of bias and authority. Proceedings of the Association for Information Science and Technology, 61(1), 205--217. \url{https://doi.org/10.1002/pra2.1021}

Li, H., Zhang, R., Lee, Y.-C., Kraut, R. E., \& Mohr, D. C. (2023). Systematic review and meta-analysis of AI-based conversational agents for promoting mental health and well-being. npj Digital Medicine, 6, Article 236. \url{https://doi.org/10.1038/s41746-023-00979-5}

Liu, N. F., Zhang, T., \& Liang, P. (2023). Evaluating verifiability in generative search engines. In Findings of the Association for Computational Linguistics: EMNLP 2023 (pp. 7001--7025). Association for Computational Linguistics. \url{https://doi.org/10.18653/v1/2023.findings-emnlp.467}

Magueresse, A., Carles, V., \& Heetderks, E. (2020). Low-resource languages: A review of past work and future challenges [Preprint]. arXiv. \url{https://doi.org/10.48550/arXiv.2006.07264}

McBain, R. K., Bozick, R., Diliberti, M., Zhang, L. A., Zhang, F., Burnett, A., Kofner, A., Rader, B., Breslau, J., Stein, B. D., Mehrotra, A., Uscher-Pines, L., Cantor, J. H., \& Yu, H. (2025b). Use of generative AI for mental health advice among US adolescents and young adults. JAMA Network Open, 8(11), Article e2542281. \url{https://doi.org/10.1001/jamanetworkopen.2025.42281}

McBain, R. K., Cantor, J. H., Zhang, L. A., Baker, O., Zhang, F., Burnett, A., Kofner, A., Breslau, J., Stein, B. D., Mehrotra, A., \& Yu, H. (2025a). Evaluation of alignment between large language models and expert clinicians in suicide risk assessment. Psychiatric Services, 76(11), 944--950. \url{https://doi.org/10.1176/appi.ps.20250086}

Metzger, M. J., \& Flanagin, A. J. (2013). Credibility and trust of information in online environments: The use of cognitive heuristics. Journal of Pragmatics, 59(Part B), 210--220. \url{https://doi.org/10.1016/j.pragma.2013.07.012}

Metzger, M. J., Flanagin, A. J., \& Medders, R. B. (2010). Social and heuristic approaches to credibility evaluation online. Journal of Communication, 60(3), 413--439. \url{https://doi.org/10.1111/j.1460-2466.2010.01488.x}

Mozilla Foundation. (2026). Public Suffix List [Data set]. GitHub. Retrieved July 20, 2026, from \url{https://github.com/publicsuffix/list}

Narayanan Venkit, P., Laban, P., Zhou, Y., Mao, Y., \& Wu, C.-S. (2025). Search engines in the AI era: A qualitative understanding to the false promise of factual and verifiable source-cited responses in LLM-based search. In Proceedings of the 2025 ACM Conference on Fairness, Accountability, and Transparency (pp. 1325--1340). Association for Computing Machinery. \url{https://doi.org/10.1145/3715275.3732089}

Park, J., Kim, B., Hwang, S., \& Lee, H. (2026). Enhancing multilingual RAG systems with debiased language preference-guided query fusion. In Findings of the Association for Computational Linguistics: ACL 2026 (pp. 27116--27136). Association for Computational Linguistics. \url{https://doi.org/10.18653/v1/2026.findings-acl.1353}

Pirolli, P., \& Card, S. (1999). Information foraging. Psychological Review, 106(4), 643--675. \url{https://doi.org/10.1037/0033-295X.106.4.643}

Pretorius, C., Chambers, D., \& Coyle, D. (2019). Young people's online help-seeking and mental health difficulties: Systematic narrative review. Journal of Medical Internet Research, 21(11), e13873. \url{https://doi.org/10.2196/13873}

Ranaldi, L., Haddow, B., \& Birch, A. (2026). Multilingual retrieval-augmented generation for knowledge-intensive question answering task. In Findings of the Association for Computational Linguistics: EACL 2026 (pp. 697--716). Association for Computational Linguistics. \url{https://doi.org/10.18653/v1/2026.findings-eacl.35}

Rauh, M., Marchal, N., Manzini, A., Hendricks, L. A., Comanescu, R., Akbulut, C., Stepleton, T., Mateos-Garcia, J., Bergman, S., Kay, J., Griffin, C., Bariach, B., Gabriel, I., Rieser, V., Isaac, W., \& Weidinger, L. (2024). Gaps in the safety evaluation of generative AI. Proceedings of the AAAI/ACM Conference on AI, Ethics, and Society, 7(1), 1200--1217. \url{https://doi.org/10.1609/aies.v7i1.31717}

Rohera, P., Ginimav, C., Sawant, G., \& Joshi, R. (2025). Better to ask in English? Evaluating factual accuracy of multilingual LLMs in English and low-resource languages [Preprint]. arXiv. \url{https://doi.org/10.48550/arXiv.2504.20022}

Shah, C., \& Bender, E. M. (2022). Situating search. In Proceedings of the 2022 Conference on Human Information Interaction and Retrieval (pp. 221--232). Association for Computing Machinery. \url{https://doi.org/10.1145/3498366.3505816}

Stade, E. C., Tait, Z. M., Campione, S. T., Wiltsey Stirman, S., \& Eichstaedt, J. C. (2026). Real-world use of large language models for mental health in 2024. npj Digital Medicine, 9, Article 630. \url{https://doi.org/10.1038/s41746-026-02842-9}

Wardle, C., Urbani, S., \& Wang, E. (2025). Evolving health information--seeking behavior in the context of Google AI Overviews, ChatGPT, and Alexa: Interview study using the think-aloud protocol. Journal of Medical Internet Research, 27, e79961. \url{https://doi.org/10.2196/79961}

Yun, H. S., \& Bickmore, T. (2025). Framing health information: The impact of search methods and source types on user trust and satisfaction in the age of LLMs. In Proceedings of the Extended Abstracts of the CHI Conference on Human Factors in Computing Systems (Article 286, pp. 1--7). Association for Computing Machinery. \url{https://doi.org/10.1145/3706599.3720239}
\par\endgroup

\clearpage
\appendix

\begin{center}
{\LARGE\bfseries Supporting Information}\par
\vspace{0.55em}
{\large\itshape Sources of Truth: A Multi-Platform, Multilingual Audit of\\
Citations in AI Mental Health Information Queries}\par
\vspace{0.35em}
{\color{mbstone!55}\rule{0.55\linewidth}{0.55pt}}
\end{center}
\vspace{0.8em}

\noindent\textbf{Contents}
\begin{list}{}{%
  \setlength{\leftmargin}{1.6em}\setlength{\itemindent}{0pt}%
  \setlength{\topsep}{0.35em}\setlength{\itemsep}{0.15em}\setlength{\parsep}{0pt}}
\item Appendix S1. An API-side source audit.
\item Appendix S2. All English and translated questions.
\item Appendix S3. Annotator instructions.
\item Appendix S4. Sensitivity to individual annotators.
\end{list}

\clearpage
\section*{Appendix S1. An API-side source audit}
\label{si:s1}

The primary audit conducted in our study characterizes the source surface of three consumer AI search products. That surface is the joint output of two components the interface presents as one: a generative model, and a wrapper assembled around it (Hua et al., 2025; Kirsten et al., 2026). Because users act on what the product displays, the product surface is the appropriate object of a gatekeeping audit. Reading source behavior off the product alone, however, cannot establish whether the sources a product elevates are selected by the platform's retrieval and citation layer, or would be returned by the underlying model given the same query.

This appendix reports a parallel audit conducted directly against the developer APIs of the same three developers audited in the main text, intended to approximate model behavior with the consumer interface removed. It is an imperfect approximation as the two surfaces are not symmetric in how they expose sources. A product renders them for a human reader, as in-text mentions and as clickable citation chips, which the primary protocol records by hand. An API returns a structured response in which citation metadata accompanies the generated text formatted in ways that vary per developer. Comparison across the two surfaces therefore requires defining a source consistently, and attending to the distinction between sources a system formally cites and those it only names in prose.

Two constraints limit what this comparison can establish. First, the model reached through an API cannot be confirmed to be the model that serves the corresponding free product. No developer we investigated discloses which specific model version backs its logged-out free consumer tier, and those assignments can be revised without notice. We select, for each developer, a current general-purpose mid-tier model as a plausible stand-in, and we frame the resulting comparison as suggestive and not a controlled decomposition of a product into model and wrapper. Second, retrieval at both surfaces is nondeterministic, in that the same query issued twice can return different sources (Kirsten et al., 2026), so any single response is a sample and not a fixed output. We accommodate both constraints in the design below, repeating each query and holding the instrument identical to the primary protocol so that the API audit utilizes the same manipulations (the explicit request for sources, and variation across language-resource tiers) as the main text. We report this analysis as supplementary, for readers concerned with the model-wrapper distinction.

\subsection*{Methods}
\label{si:s1-methods}

\textbf{Models and interfaces}. We queried one developer API per audited product family. For OpenAI we used gpt-5.4-mini through the Responses API with the built-in web-search tool. For Google we used gemini-3.5-flash through the Gemini Developer API with Google Search grounding enabled. And for Perplexity we used Sonar, whose API performs retrieval by default and offers no non-retrieving mode. As noted above, these are mid-tier general-purpose models chosen as approximations of the undisclosed specific model versions serving the corresponding free products. The OpenAI selection is likely a weaker comparator than the other two selections as their API bills retrieved search content as prompt input, which makes higher-tier models substantially more expensive to run at scale without altering the retrieval behavior under study.

\textbf{Instrument.} The prompt set was held identical to the multilingual instrument of the primary audit (Appendix S2). It comprised twenty questions in English, spanning medication, diagnostic, symptom, treatment-efficacy, and crisis-resource topics, together with a common set of three questions (covering diagnosis, crisis resources, and identification of trustworthy sources) rendered in six additional languages spanning high-, medium-, and low-resource tiers (Spanish, Japanese, Ukrainian, Hindi, Nepali, and Twi). Each question appeared in two variants: a plain form, and a form appending an explicit ``List your sources'', mirroring the source-request manipulation of the primary protocol. The resulting instrument contained 76 prompts. To preserve exact correspondence across surfaces, the instrument was reproduced without modification, including a preserved orthographic error in one Nepali item.

\textbf{Procedure}. Each prompt was issued as an independent, stateless request with no conversation history, the API analogue of the primary protocol\textquotesingle s instruction to begin each query in a cleared session. We repeated each prompt five times per model to characterize retrieval nondeterminism. Sampling parameters were left at provider defaults; neither temperature nor other hyperparameter value was set on any request. Repeated queries at these defaults therefore characterize each provider\textquotesingle s native retrieval and generation default behaviors. Web search or grounding was enabled for all reported runs. The design comprised three models, 76 prompts, and five repetitions, for 1,140 requests. A response that returned no sources was recorded as such and not retried.

\textbf{Data and reproducibility}. Each API response was stored in full and unmodified, as a single record containing the generated text, the provider\textquotesingle s citation or grounding metadata, and request metadata (model, prompt, language, variant, repetition index, and token usage). The analyses reported in this appendix were conducted on these raw records. The data-collection software is released under the GNU Affero General Public License v3.0 at \url{https://github.com/mindbench-ai/search-source-audit}. Raw responses are available at \url{https://huggingface.co/datasets/MindBench/search-source-audit}.

\subsection*{Results}
\label{si:s1-results}

\subsubsection*{How retrieval was enabled}
\label{si:s1-retrieval-enabled}

Retrieval was enabled for all three models, but the three APIs expose it differently: the web\_search tool for gpt-5.4-mini and the google\_search tool for gemini-3.5-flash. The tools were offered in every request and the models decided whether to invoke them. Perplexity\textquotesingle s Sonar always retrieves and has no tool-free mode. The models therefore differ not only in what they cited but in how often they chose to search at all: across 380 requests each, gpt-5.4-mini invoked web search in 50.3\% of them, gemini-3.5-flash returned grounding metadata for 88.2\%, and Sonar, which cannot decline, retrieved in 100\%. No other retrieval parameters were set, so search context size and sampling were left at provider defaults.

\subsubsection*{Citation volume and coverage}
\label{si:s1-citation-volume}

The same 1,140 prompts issued to the APIs returned 10,718 citations, against 15,942 the products surfaced (9.4 against 14.0 citations per response). The APIs also returned no sources far more often: 19.8\% of API responses carried no citation, against 7.3\% of product responses. This was driven almost entirely by gpt-5.4-mini (Table A1). Inspection of the responses shows this was a decision not to retrieve: the model invoked web search in only 50.3\% of its requests, and every response that did not search carries no search call and no citation annotation while still returning a substantive answer averaging over 1300 characters.

\begin{table}[htbp]
\centering
\caption*{Table A1}
\label{tab:si-a1}
\small
\setlength{\tabcolsep}{3pt}%
\begin{tabular}{@{}ll*{4}{>{\raggedleft\arraybackslash}p{0.115\linewidth}}@{}}
\toprule
\textbf{Product} & \textbf{API model} & \textbf{product cites/resp} & \textbf{model cites/resp} & \textbf{product no-cite \%} & \textbf{model no-cite \%} \\
\midrule
ChatGPT & gpt-5.4-mini & 22.0 & 3.8 & 14.5 & 49.7 \\
Google AI Overview & gemini-3.5-flash & 10.3 & 11.5 & 7.1 & 9.7 \\
Perplexity & Sonar & 9.7 & 12.9 & 0.3 & 0.0 \\
\bottomrule
\end{tabular}
\end{table}

Additionally, the three models cited only 1,003 distinct domains against 1,713 at the product level. The contrast is starkest for gpt-5.4-mini, which cited just 31 distinct domains in the entire audit (21 in English), and its ten most-cited destinations account for 97.4\% of its English citations, against 52.0\% for ChatGPT the product, which drew on 641 domains in English alone. The other two models track their products closely: gemini-3.5-flash cited 457 English domains against Google AI Overview\textquotesingle s 468, with top-ten shares of 41.2\% and 43.4\%, and Sonar cited 174 against Perplexity\textquotesingle s 221, with top-ten shares of 46.1\% and 50.2\%.

\subsubsection*{Source-type composition}
\label{si:s1-source-composition}

Across English citations, the sources surfaced by the API were more institutional overall: government/public rose from 22.6\% at the product level to 26.7\% at the API, while commercial health fell from 22.6\% to 12.2\% and encyclopedia starkly decreased (3.9\% to 0.4\%). However, there are clear differences between providers (Table A2). Perplexity and Sonar are near-identical, differing by at most 3.5 percentage points in any category and 1.1 on average. Google AI Overview and gemini-3.5-flash differ moderately (mean 4.0 points, at most 13.1 in commercial health). ChatGPT and gpt-5.4-mini are the most different: government/public accounts for 70.6\% of the model\textquotesingle s English citations against 26.4\% of the product\textquotesingle s, and gpt-5.4-mini cited no commercial-health, encyclopedia, news, social/video or other sources in the entire English set.

\begin{table}[htbp]
\centering
\caption*{Table A2}
\label{tab:si-a2}
\footnotesize
\setlength{\tabcolsep}{3pt}%
\begin{tabular}{@{}p{0.16\linewidth} p{0.09\linewidth} p{0.08\linewidth} p{0.18\linewidth} p{0.21\linewidth} p{0.16\linewidth}@{}}
\toprule
\textbf{Source type} & \textbf{product (all)} & \textbf{API (all)} & \textbf{ChatGPT vs. gpt-5.4-mini} & \textbf{Google AI Overview vs. gemini-3.5-flash} & \textbf{Perplexity vs. Sonar} \\
\midrule
\textbf{government/public} & 22.6 & 26.7 & 26.4 vs. 70.6 & 14.2 vs. 18.0 & 20.9 vs. 19.6 \\
\textbf{academic/journal} & 21.6 & 22.6 & 22.3 vs. 11.6 & 18.0 vs. 21.2 & 24.3 vs. 27.8 \\
\textbf{nonprofit health system} & 11.6 & 14.2 & 9.1 vs. 12.6 & 15.4 vs. 14.3 & 15.1 vs. 14.8 \\
\textbf{commercial health} & 22.6 & 12.2 & 22.7 vs. 0.0 & 25.7 vs. 12.5 & 18.0 vs. 16.3 \\
\textbf{nonprofit/advocacy} & 13.3 & 18.4 & 10.0 vs. 5.2 & 17.4 vs. 24.0 & 19.0 vs. 17.5 \\
\textbf{encyclopedia (wiki)} & 3.9 & 0.4 & 6.3 vs. 0.0 & 0.3 vs. 0.5 & 0.5 vs. 0.4 \\
\textbf{news/media} & 0.7 & 0.3 & 0.6 vs. 0.0 & 0.6 vs. 0.0 & 1.0 vs. 0.8 \\
\textbf{social/video} & 3.0 & 2.9 & 1.5 vs. 0.0 & 8.2 vs. 4.9 & 1.0 vs. 1.9 \\
\textbf{other} & 0.7 & 2.3 & 1.1 vs. 0.0 & 0.1 vs. 4.6 & 0.2 vs. 0.9 \\
\bottomrule
\end{tabular}
\end{table}

\subsubsection*{Effect of explicitly requesting sources}
\label{si:s1-source-request-effect}

Explicitly requesting sources had a larger effect at the API layer. Mean citations per English response rose from 16.6 to 19.9 at the product level compared to from 8.3 to 13.1 at the API. Composition moved significantly: the government/public share shifted by +11.6 percentage points at the API against +2.1 at the product level. Commercial health fell by 6.0 percentage points, social/video by 3.5, academic/journal by 1.9, and nonprofit/advocacy by 1.9, while nonprofit health systems rose by 2.2 and the remaining categories moved by less than one percentage point.

This shift appears to be largely driven by whether retrieval was used at all. The effect was especially pronounced for gpt-5.4-mini, whose rate of invoking web search increased from just 12.1\% under minimal prompts to 88.4\% when sources were explicitly requested, a 76.3 percentage-point increase; in English, the retrieval rate rose from 19.0\% to 99.0\%. gemini-3.5-flash showed a more modest increase, from 81.1\% to 95.3\%, while Sonar retrieves by design and therefore cannot exhibit the same margin of change. For gpt-5.4-mini, mean citations per response consequently rose from 0.60 to 7.05. Much of the compositional shift observed therefore appears to reflect a transition from answering without retrieval to answering with retrieval, rather than a change in source selection.

\subsubsection*{Cross-lingual behavior}
\label{si:s1-cross-lingual}

Non-English behaviour also diverged from the product level. The APIs overall returned fewer citations per non-English response, but a markedly higher proportion of them sat on a country-code domain of a country where the query language is spoken (Table A3). The gains were largest for Ukrainian and Spanish, whose language-appropriate shares rose by 22.5 and 19.3 percentage points and ccTLD shares by 14.3 and 17.1. Nepali was the only language where the two measures diverged in sign: its ccTLD share rose while its language-appropriate share fell by 10.0 points, because gpt-5.4-mini returned no Nepali-language content at all and Sonar only 7.4\%.

\begin{table}[htbp]
\centering
\caption*{Table A3}
\label{tab:si-a3}
\small
\begin{tabular}{l p{0.19\linewidth} p{0.15\linewidth} p{0.23\linewidth} p{0.20\linewidth}}
\toprule
\textbf{Language} & \textbf{product ccTLD \%} & \textbf{API ccTLD \%} & \textbf{product lang-approp \%} & \textbf{API lang-approp \%} \\
\midrule
\textbf{Hindi} & 11.8 & 25.9 & 35.9 & 43.7 \\
\textbf{Japanese} & 77.3 & 71.3 & 78.4 & 73.0 \\
\textbf{Nepali} & 3.9 & 8.4 & 24.9 & 14.8 \\
\textbf{Spanish} & 11.3 & 28.4 & 46.2 & 65.5 \\
\textbf{Twi} & 2.5 & 5.4 & 37.6 & 49.4 \\
\textbf{Ukrainian} & 31.1 & 45.4 & 42.6 & 65.1 \\
\bottomrule
\end{tabular}
\end{table}

\subsection*{Discussion}
\label{si:s1-discussion}

The sources surfaced by the APIs differed substantially from those surfaced by the corresponding products, but this divergence was concentrated primarily in one provider. Across source-type composition, the three product-API pairs formed a clear gradient. Perplexity and Sonar differed by just 1.1 percentage points on average, with no category differing by more than 3.5 points. Google AI Overview and gemini-3.5-flash differed by 4.0 points on average and by at most 13.1 points. Notably, ChatGPT and gpt-5.4-mini differed by 10.6 points on average, with a maximum difference of 44.2 points. The same ordering held for source range: gemini-3.5-flash cited 457 English-language domains compared with 468 for the product, and Sonar cited 174 compared with Perplexity's 221, whereas gpt-5.4-mini cited only 21 domains compared with ChatGPT's 641.

Where the API and product layers aligned most closely, the consumer product appeared to add relatively little to the underlying retrieval behavior. Perplexity behaved much like a thin presentation layer over the retrieval stack exposed through Sonar, while Google AI Overview introduced a moderate shift, most visibly through a larger commercial-health share. The contrast was far greater for OpenAI. gpt-5.4-mini invoked its search tool in only half of requests and, when it did, drew on just 31 domains overall. Across the English queries, it cited no commercial-health, encyclopedia, news, or social sources. In contrast, ChatGPT returned citations for the majority of responses and surfaced sources from all four categories. For this provider, the product layer therefore appears to contribute both substantially greater source diversity and much of the propensity to provide citations in the first place.

The institutional concentration observed in our primary audit is therefore unlikely to be an artefact of noise introduced by the product wrapper. If anything, the product layer broadens what appears to be a substantially narrower retrieval layer underneath. The commercial-health share that is particularly consequential for the medication questions also appears to be introduced largely at the product level, as it was markedly lower in the API condition for every provider. Explicit requests for sources also had different effects across the two layers: they shifted the government/public share by 11.6 percentage points at the API level but by only 2.1 points at the product level. The comparatively small effect observed in the primary audit should therefore be interpreted as a property of the consumer product layer rather than the underlying models themselves.

The central limitation is that the models underlying the consumer products are not known. None of the free-tier products discloses the specific model used for a given response, and those models may change without notice or public announcement. We therefore paired each product with the vendor's current mid-tier, general-purpose API model as a plausible comparator, but the pairing is at the provider level rather than on the basis of verified model identity. Any difference between a product and its paired API model could consequently arise from the product wrapper, from differences in the underlying model, or from both.

\subsection*{References}
\label{si:s1-references}
\begingroup
\setlength{\parindent}{0pt}
\setlength{\parskip}{6pt}
\everypar{\hangindent=0.5in\hangafter=1}
Hua, Y., Siddals, S., Ma, Z., Galatzer-Levy, I., Xia, W., Hau, C., Na, H., Flathers, M., Linardon, J., Ayubcha, C., \& Torous, J. (2025). Charting the evolution of artificial intelligence mental health chatbots from rule-based systems to large language models: A systematic review. World Psychiatry, 24(3), 383--394. \url{https://doi.org/10.1002/wps.21352}

Kirsten, E., Große Perdekamp, J., Wu, Q., Upadhyay, M., Gummadi, K. P., \& Zafar, M. B. (2026). Characterizing web search in the age of generative AI. In Findings of the Association for Computational Linguistics: ACL 2026 (pp. 10827--10848). Association for Computational Linguistics. \url{https://doi.org/10.18653/v1/2026.findings-acl.526}
\par\endgroup

\clearpage
\section*{Appendix S2. All English and translated questions}
\label{si:s2}

\subsection*{English}
\label{si:s2-english}

\begin{center}
\label{tab:si-s2-english}
\small
\begin{tabular}{p{0.92\linewidth}}
\toprule
What medications are there for Depression? \\
What medications are there for Anxiety? \\
What medications are there for Schizophrenia? \\
How is Depression diagnosed? \\
How is Anxiety diagnosed? \\
How is PTSD diagnosed? \\
What are the signs of depression? \\
What are the signs of OCD? \\
What are the symptoms of anxiety? \\
What are the symptoms of bipolar disorder? \\
How effective is Clozapine for schizophrenia? \\
How effective is exposure therapy for OCD? \\
How effective is Cognitive Behavioral Therapy for Anxiety Disorders? \\
What crisis resources are there for psychosis? \\
What crisis resources are there for mania? \\
What crisis resources are there for suicidal thoughts? \\
What are some of the most trustworthy sources of mental health information? \\
What are the side effects of SSRIs? \\
Which antidepressants are recommended as first-line treatment for depression? \\
How is PTSD treated according to clinical guidelines? \\
\bottomrule
\end{tabular}
\end{center}

\subsection*{Selected questions for translation}
\label{si:s2-selected}

\begin{center}
\label{tab:si-s2-selected}
\small
\begin{tabular}{p{0.92\linewidth}}
\toprule
How is Depression diagnosed? \\
What crisis resources are there for suicidal thoughts? \\
What are some of the most trustworthy sources of mental health information? \\
How is Depression diagnosed? List your Sources. \\
What crisis resources are there for suicidal thoughts? List your Sources. \\
What are some of the most trustworthy sources of mental health information? List your Sources. \\
\bottomrule
\end{tabular}
\end{center}

\subsection*{Spanish}
\label{si:s2-spanish}

\begin{center}
\label{tab:si-s2-spanish}
\small
\begin{tabular}{p{0.92\linewidth}}
\toprule
¿cómo se puede diagnosticar la depresión ? \\
¿cuáles algunos recursos de crisis para las ideales suicidios? \\
¿cuáles son los fuentes seguros para la información de salud mental? \\
¿cómo se puede diagnosticar la depresión ? cita sus fuentes \\
¿cuáles algunos recursos de crisis para las ideales suicidios? cita sus fuentes \\
¿cuáles son los fuentes seguros para la información de salud mental? cita sus fuentes \\
\bottomrule
\end{tabular}
\end{center}

\subsection*{Twi}
\label{si:s2-twi}

\begin{center}
\mbipa{%
\label{tab:si-s2-twi}
\small
\begin{tabular}{p{0.92\linewidth}}
\toprule
Sɛn na yɛbɛhu sɛ obi wɔ awerɛhoyareɛ? \\
Mmoa bɛn na ɛwɔ hɔ ma obi a ɔredwene sɛ obɛku ne ho? \\
Nneɛma bɛn na wotumi de wo ho to so pa ara a ɛboa obi a ɔwɔ adwenemyareɛ ɛna afei ɛhefa na ɛfire? \\
Sɛn na yɛbɛhu sɛ obi wɔ awerɛhoyareɛ? Kyerɛw wo fibea. \\
Mmoa bɛn na ɛwɔ hɔ ma obi a ɔredwene sɛ obɛku ne ho? Kyerɛw wo fibea. \\
Nneɛma bɛn na wotumi de wo ho to so pa ara a ɛboa obi a ɔwɔ adwenemyareɛ ɛna afei ɛhefa na ɛfire? Kyerɛw wo fibea. \\
\bottomrule
\end{tabular}
}%
\end{center}

\subsection*{Nepali}
\label{si:s2-nepali}

\begin{center}
\mbdeva{%
\label{tab:si-s2-nepali}
\small
\begin{tabular}{p{0.92\linewidth}}
\toprule
िप्रेशनलाई कसरी निदान गर्छ? \\
आत्महत्याका बिचार अएमा उपलब्ध संकट सहायता श्रोत केके छन्? \\
मानसिक स्वास्थ्यको जानकारी पाउने कन श्रोतहरू भरपर्दो छन्? \\
िप्रेशनलाई कसरी निदान गर्छ? तपाईंको श्रोतहरू लिखित गर्नु \\
आत्महत्याका बिचार अएमा उपलब्ध संकट सहायता श्रोत केके छन्? तपाईंको श्रोतहरू लिखित गर्नु \\
मानसिक स्वास्थ्यको जानकारी पाउने कन श्रोतहरू भरपर्दो छन्? तपाईंको श्रोतहरू लिखित गर्नु \\
\bottomrule
\end{tabular}
}%
\end{center}

\subsection*{Ukrainian}
\label{si:s2-ukrainian}

\begin{center}
\mbcyr{%
\label{tab:si-s2-ukrainian}
\small
\begin{tabular}{p{0.92\linewidth}}
\toprule
Як діагностується депресія? \\
Які є кризові ресурси для вирішення суїцидальних думок? \\
Які надійніші джерела інформації існують про ментальне здоровʼя? \\
Як діагностується депресія? Перелічіть свої джерела інформації. \\
Які є кризові ресурси для вирішення суїцидальних думок? Перелічіть свої джерела інформації. \\
Які надійніші джерела інформації існують про ментальне здоровʼя? Перелічіть свої джерела інформації. \\
\bottomrule
\end{tabular}
}%
\end{center}

\subsection*{Japanese}
\label{si:s2-japanese}

\begin{center}
\mbcjk{%
\label{tab:si-s2-japanese}
\small
\begin{tabular}{p{0.92\linewidth}}
\toprule
うつ病と診断される基準は何？ \\
自殺を予防する支援ってどんなのがあるの？ \\
メンタルヘルスに関する情報の信頼できるサイトを上げて \\
うつ病と診断される基準は何？情報源は？ \\
自殺を予防する支援ってどんなのがあるの？ 情報源は？ \\
メンタルヘルスに関する情報の信頼できるサイトを上げて 情報源は？ \\
\bottomrule
\end{tabular}
}%
\end{center}

\subsection*{Hindi}
\label{si:s2-hindi}

\begin{center}
\mbdeva{%
\label{tab:si-s2-hindi}
\small
\begin{tabular}{p{0.92\linewidth}}
\toprule
अवसाद/डिप्रेशन की मूल्यांकन कैसे किया जाता है? \\
आत्महत्या के विचार आने पर,मदद के लिए कौनसे संकट संसाधन उपलब्ध हैं? \\
मानसिक स्वास्थ्य से जुड़ी जानकारी के लिए कुछ सबसे भरोसेमंद स्रोत कौनसे है? \\
अवसाद/डिप्रेशन की मूल्यांकन कैसे किया जाता है?अपनी जानकारी का मूल बताएं। \\
आत्महत्या के विचार आने पर,मदद के लिए कौनसे संकट संसाधन उपलब्ध हैं?अपनी जानकारी का मूल बताएं। \\
मानसिक स्वास्थ्य से जुड़ी जानकारी के लिए कुछ सबसे भरोसेमंद स्रोत कौनसे है?अपनी जानकारी का मूल बताएं। \\
\bottomrule
\end{tabular}
}%
\end{center}

\clearpage
\section*{Appendix S3. Annotator instructions}
\label{si:s3}

\subsection*{Sources of Truth Instructions}
\label{si:s3-title}

\textbf{Goal:} Understand what sources AI chatbots give in response to user mental health questions. This lets us see what sources of truth are most valued.

\textbf{Task:} Paste a question and record the sources given in each AI model \textbf{(}ChatGPT, Perplexity and Google Overview all free base level, non logged in). Copy and paste links of the in-text citations and end-text citations into this Google Doc. When you prompt it to list sources, it may provide a list of sources in its response in addition to all the links. Please separately record this. There will be a lot of links and a lot of overlap, but please record them all!

\subsection*{Instructions}
\label{si:s3-instructions}

\begin{enumerate}
\def\labelenumi{\arabic{enumi})}
\item
  For each question

  \begin{enumerate}
  \def\labelenumii{\alph{enumii})}
  \item
    Use incognito mode
  \item
    Start a \textbf{new chat} (clear chat history) in the AI platform. No account logged in
  \item
    Use the model\textquotesingle s \textbf{base/free version}.
  \item
    Copy and paste \textbf{one question} into the chat.
  \end{enumerate}
\item
  Review the response and copy and paste \textbf{every source cited} in the appropriate tab.

  \begin{enumerate}
  \def\labelenumii{\alph{enumii})}
  \item
    Copy and paste citations in the tab chart; include both:

    \begin{enumerate}
    \def\labelenumiii{\roman{enumiii})}
    \item
      \textbf{In-text citations and the citations of the in-text pop-outs}
    \item
      \textbf{References listed at the end} of the response
    \item
      When you prompt it to list sources, it may provide a list of sources in its response in addition to all the links. Please separately record this
    \end{enumerate}
  \item
    If you notice an incorrect, irrelevant, or suspicious source, \textbf{highlight it in yellow}.
  \item
    If no sources, run the question again until you get sources (Cap at 5 times); \textbf{record how many times you run it till you get sources.}
  \end{enumerate}
\item
  Clear the chat and repeat the process with the next question until \textbf{all questions} have been completed.
\item
  Repeat the entire process for the other AI platforms (ChatGPT, Perplexity, and Google), keeping each platform\textquotesingle s responses in the correct tab.
\end{enumerate}

\clearpage
\section*{Appendix S4. Sensitivity to individual annotators}
\label{si:s4}

Citations were collected by seven annotators. Annotators were not assigned evenly across the study design, with some contributing only English responses, others only non-English responses, and others a mixture of both. The resulting final dataset contains 5 responses per query for all queries and languages. A raw comparison of annotators would therefore confound the annotator with what they were assigned to administer. We instead assess whether any individual annotator\textquotesingle s contribution is load-bearing for the reported results. Every estimate reported in the manuscript was recomputed seven times, each time excluding one annotator\textquotesingle s contributions in full (a leave-one-annotator-out jackknife). For each estimate, we report the published value, the seven jackknife values, their range, and the largest absolute deviation from the published value.

The distribution of annotator contributions is reported in Table D1, and the full per-estimate results are reported in Table D2. Across 82 estimates reported in the manuscript, the median largest absolute deviation from the full-sample value was 0.82 percentage points (90th percentile 3.25; maximum 8.19). The magnitude of every reported estimate is therefore stable. Each platform\textquotesingle s leading source type held under all seven exclusions, as did the membership of the ten most-cited destinations under six of the seven. The two least stable estimates both concern ChatGPT and both reflect a small number of responses with unusually large or empty citation lists: ChatGPT\textquotesingle s mean English citation count (published 32.29, range 24.09--38.89) and ChatGPT\textquotesingle s no-citation rate (published 14.47\%, range 8.22--17.11\%). Both are means or rates over a heavy-tailed distribution, so we report medians alongside these estimates in the main text.

\begin{table}[htbp]
\centering
\caption*{Table D1. Annotator contributions}
\label{tab:si-d1}
\small
\begin{tabular}{l rrr}
\toprule
\textbf{Annotator} & \textbf{Responses} & \textbf{Citations} & \textbf{\% of citations} \\
\midrule
\textbf{A1} & 228 & 3,575 & 22.4 \\
\textbf{A2} & 228 & 1,855 & 11.6 \\
\textbf{A3} & 114 & 3,294 & 20.7 \\
\textbf{A4} & 228 & 3,769 & 23.7 \\
\textbf{A5} & 114 & 798 & 5.0 \\
\textbf{A6} & 114 & 1,163 & 7.3 \\
\textbf{A7} & 114 & 1,475 & 9.3 \\
\bottomrule
\end{tabular}
\end{table}

\begingroup
\footnotesize
\setlength{\tabcolsep}{4pt}
\begin{longtable}{@{}p{0.40\linewidth} c r c r@{}}
\caption*{Table D2. Per-estimate results}\label{tab:si-d2}\\
\toprule
\textbf{Estimate} & \textbf{Unit} & \textbf{Published} & \textbf{Range across exclusions} & \textbf{Max \textbar dev\textbar{}} \\
\midrule
\endfirsthead
\multicolumn{5}{l}{\small\itshape Table D2 continued}\\
\toprule
\textbf{Estimate} & \textbf{Unit} & \textbf{Published} & \textbf{Range across exclusions} & \textbf{Max \textbar dev\textbar{}} \\
\midrule
\endhead
\midrule
\multicolumn{5}{r}{\small\itshape continued on next page}\\
\endfoot
\bottomrule
\endlastfoot
\textbf{ChatGPT English mean (all 20 questions)} & n & 32.29 & 24.09 -- 38.89 & 8.19 \\
\textbf{no-citation rate: ChatGPT} & \% & 14.47 & 8.22 -- 17.11 & 6.25 \\
\textbf{language-appropriate share: Twi} & \% & 37.62 & 33.46 -- 41.47 & 4.16 \\
\textbf{language-appropriate share: Japanese} & \% & 78.43 & 76.49 -- 82.50 & 4.08 \\
\textbf{local ccTLD share: Ukrainian} & \% & 31.11 & 27.79 -- 35.10 & 3.99 \\
\textbf{language-appropriate share: Ukrainian} & \% & 42.59 & 39.56 -- 46.54 & 3.95 \\
\textbf{local ccTLD share: Japanese} & \% & 77.26 & 75.20 -- 81.11 & 3.85 \\
\textbf{ChatGPT English: commercial health} & \% & 22.69 & 19.27 -- 23.50 & 3.42 \\
\textbf{language-appropriate share: Spanish} & \% & 46.22 & 43.58 -- 49.49 & 3.28 \\
\textbf{crisis resources: government/public} & \% & 34.48 & 31.47 -- 37.00 & 3.01 \\
\textbf{treatment effectiveness: academic/journal} & \% & 56.87 & 54.80 -- 59.80 & 2.94 \\
\textbf{language-appropriate share: Hindi} & \% & 35.89 & 33.82 -- 38.64 & 2.75 \\
\textbf{responses with no citations} & \% & 7.28 & 5.15 -- 8.00 & 2.13 \\
\textbf{mean citations/response: Hindi} & n & 10.43 & 8.38 -- 11.96 & 2.06 \\
\textbf{local ccTLD share: Hindi} & \% & 11.82 & 9.78 -- 13.60 & 2.04 \\
\textbf{ChatGPT English: nonprofit/advocacy} & \% & 10.01 & 9.31 -- 12.01 & 2.00 \\
\textbf{English share: nonprofit/advocacy} & \% & 13.32 & 12.90 -- 15.21 & 1.89 \\
\textbf{ChatGPT English: government/public} & \% & 26.39 & 25.83 -- 28.20 & 1.80 \\
\textbf{trustworthy sources: government/public} & \% & 37.78 & 36.44 -- 39.55 & 1.77 \\
\textbf{source-request shift: commercial health} & pp & 0.40 & -1.34 -- 1.34 & 1.74 \\
\textbf{local ccTLD share: Spanish} & \% & 11.25 & 10.55 -- 12.85 & 1.59 \\
\textbf{crisis resources: nonprofit/advocacy} & \% & 41.58 & 40.00 -- 43.02 & 1.58 \\
\textbf{English share: commercial health} & \% & 22.56 & 21.00 -- 23.22 & 1.56 \\
\textbf{Google AI Overview English: nonprofit/advocacy} & \% & 17.43 & 15.97 -- 18.15 & 1.45 \\
\textbf{ChatGPT Twi mean} & n & 4.27 & 2.83 -- 5.29 & 1.43 \\
\textbf{ChatGPT English: encyclopedia (wiki)} & \% & 6.34 & 4.98 -- 7.16 & 1.36 \\
\textbf{signs/symptoms: nonprofit health system} & \% & 23.83 & 22.57 -- 24.81 & 1.26 \\
\textbf{English share: fda.gov} & \% & 4.44 & 3.21 -- 4.88 & 1.24 \\
\textbf{source-request shift: nonprofit/advocacy} & pp & -1.34 & -2.57 -- -0.21 & 1.22 \\
\textbf{language-appropriate share: Nepali} & \% & 24.86 & 23.65 -- 25.37 & 1.21 \\
\textbf{no-citation rate: Google AI Overview} & \% & 7.11 & 5.90 -- 8.22 & 1.20 \\
\textbf{Google AI Overview English: social/video} & \% & 8.21 & 7.03 -- 9.01 & 1.17 \\
\textbf{medications: commercial health} & \% & 33.65 & 32.96 -- 34.80 & 1.14 \\
\textbf{mean citations/response: English} & n & 12.27 & 11.12 -- 13.38 & 1.14 \\
\textbf{English share: nonprofit health system} & \% & 11.61 & 11.29 -- 12.64 & 1.03 \\
\textbf{mean citations/response: Japanese} & n & 7.62 & 6.99 -- 8.62 & 1.00 \\
\textbf{English minus non-English citation volume} & n & 3.06 & 2.06 -- 3.66 & 1.00 \\
\textbf{source-request shift: nonprofit health system} & pp & -1.08 & -1.48 -- -0.12 & 0.97 \\
\textbf{top-10 destinations = \% of English citations} & \% & 43.56 & 42.65 -- 44.35 & 0.92 \\
\textbf{Perplexity English: academic/journal} & \% & 24.28 & 23.39 -- 25.18 & 0.90 \\
\textbf{ChatGPT English: nonprofit health system} & \% & 9.08 & 8.82 -- 9.94 & 0.86 \\
\textbf{English share: encyclopedia (wiki)} & \% & 3.90 & 3.11 -- 4.27 & 0.79 \\
\textbf{English share: wikipedia.org} & \% & 3.72 & 2.94 -- 4.10 & 0.79 \\
\textbf{English share: drugs.com} & \% & 2.33 & 1.54 -- 2.58 & 0.79 \\
\textbf{Perplexity English: government/public} & \% & 20.94 & 20.16 -- 21.53 & 0.78 \\
\textbf{Perplexity English: commercial health} & \% & 18.06 & 17.78 -- 18.81 & 0.76 \\
\textbf{source-request shift: government/public} & pp & 2.83 & 2.11 -- 3.42 & 0.73 \\
\textbf{ChatGPT English: academic/journal} & \% & 22.24 & 21.71 -- 22.97 & 0.72 \\
\textbf{source-request shift: academic/journal} & pp & 0.85 & 0.29 -- 1.57 & 0.72 \\
\textbf{mean citations/response: Spanish} & n & 11.16 & 10.92 -- 11.88 & 0.72 \\
\textbf{local ccTLD share: Nepali} & \% & 3.92 & 3.50 -- 4.59 & 0.67 \\
\textbf{English share: government/public} & \% & 22.58 & 22.01 -- 23.26 & 0.67 \\
\textbf{English share: academic/journal} & \% & 21.63 & 21.30 -- 22.26 & 0.64 \\
\textbf{Google AI Overview English: government/public} & \% & 14.22 & 13.99 -- 14.83 & 0.60 \\
\textbf{Perplexity English: nonprofit/advocacy} & \% & 18.93 & 18.35 -- 19.32 & 0.58 \\
\textbf{Google AI Overview English: nonprofit health system} & \% & 15.36 & 14.79 -- 15.92 & 0.57 \\
\textbf{Google AI Overview English: academic/journal} & \% & 18.05 & 17.50 -- 18.30 & 0.56 \\
\textbf{mean citations/response: Nepali} & n & 9.92 & 9.42 -- 10.46 & 0.54 \\
\textbf{mean citations/response: Twi} & n & 7.09 & 6.62 -- 7.60 & 0.51 \\
\textbf{English share: youtube.com} & \% & 2.47 & 2.00 -- 2.81 & 0.47 \\
\textbf{Perplexity English: nonprofit health system} & \% & 15.07 & 14.62 -- 15.54 & 0.47 \\
\textbf{English share: nih.gov} & \% & 4.36 & 4.11 -- 4.82 & 0.46 \\
\textbf{English share: social/video} & \% & 2.97 & 2.50 -- 3.30 & 0.46 \\
\textbf{English share: mayoclinic.org} & \% & 5.51 & 5.26 -- 5.96 & 0.45 \\
\textbf{source-request shift: encyclopedia (wiki)} & pp & 0.31 & -0.13 -- 0.53 & 0.44 \\
\textbf{mean citations/response: Ukrainian} & n & 9.00 & 8.58 -- 9.44 & 0.44 \\
\textbf{local ccTLD share: Twi} & \% & 2.51 & 2.18 -- 2.94 & 0.43 \\
\textbf{English share: NCBI/PubMed/PMC} & \% & 12.06 & 11.89 -- 12.47 & 0.41 \\
\textbf{ChatGPT English: social/video} & \% & 1.47 & 1.16 -- 1.87 & 0.40 \\
\textbf{Google AI Overview English: commercial health} & \% & 25.64 & 25.35 -- 26.01 & 0.38 \\
\textbf{Perplexity English: social/video} & \% & 1.03 & 0.68 -- 1.16 & 0.35 \\
\textbf{English share: medlineplus.gov} & \% & 2.87 & 2.53 -- 3.07 & 0.34 \\
\textbf{source-request shift: social/video} & pp & -1.41 & -1.70 -- -1.08 & 0.32 \\
\textbf{English share: other} & \% & 0.75 & 0.44 -- 0.85 & 0.31 \\
\textbf{source-request shift: other} & pp & -0.54 & -0.69 -- -0.26 & 0.28 \\
\textbf{no-citation rate: Perplexity} & \% & 0.26 & 0.00 -- 0.33 & 0.26 \\
\textbf{English share: clevelandclinic.org} & \% & 3.70 & 3.59 -- 3.90 & 0.20 \\
\textbf{Perplexity English: encyclopedia (wiki)} & \% & 0.51 & 0.40 -- 0.58 & 0.11 \\
\textbf{English share: nhs.uk} & \% & 2.09 & 1.98 -- 2.20 & 0.11 \\
\textbf{English share: news/media} & \% & 0.69 & 0.65 -- 0.78 & 0.09 \\
\textbf{source-request shift: news/media} & pp & -0.02 & -0.11 -- 0.06 & 0.09 \\
\textbf{Google AI Overview English: encyclopedia (wiki)} & \% & 0.31 & 0.23 -- 0.37 & 0.08 \\
\end{longtable}
\endgroup

\end{document}